\documentclass[prc,nofootinbib,showpacs,twocolumn]{revtex4}

\usepackage{graphicx}
\usepackage[usenames,dvipsnames]{color}
\usepackage{amsmath,amssymb,bbold}
\usepackage{hyperref}

\begin{document}

\title{A Novel Mechanism for $J/\psi$ Disintegration in 
Relativistic Heavy Ion Collisions}

\author{Abhishek Atreya}
\email{atreya@iopb.res.in}
\affiliation{Institute of Physics, Bhubaneswar, Odisha, India 751005}
\author{Partha Bagchi}
\email{partha@iopb.res.in}
\affiliation{Institute of Physics, Bhubaneswar, Odisha, India 751005}
\author{Ajit M. Srivastava}
\email{ajit@iopb.res.in}
\affiliation{Institute of Physics, Bhubaneswar, Odisha, India 751005}

\begin{abstract}
In this paper we discuss the possibility of $J/\psi$ disintegration
due to  $Z(3)$ domain walls that are expected to form in a QGP
medium. These domain walls give rise to localized color electric field
which disintegrates $J/\psi$, on interaction, by changing its color
composition and simultaneously exciting it to higher states of
$c\bar{c}$ system.
\end{abstract}

\pacs{PACS numbers: 25.75.-q, 11.27.+d, 14.40.Lb, 12.38.Mh}
\maketitle

\section{Introduction}
The ongoing relativistic heavy ion collision experiments at RHIC (BNL)
and LHC (CERN) have provided very valuable insights in understanding
certain aspects of QCD. One such aspect is the existence of a new phase of
matter known as quark-gluon plasma (QGP). QGP is essentially the 
deconfined phase of QCD, where free quarks and gluons exist in thermal
equilibrium. Matsui and Satz \cite{Matsui:1986dk} proposed that due
to the presence of this medium, potential between $q\bar{q}$ is Debye
screened, resulting in the swelling of quarkonia. If the Debye
screening length of the medium is less than the radius of quarkonia,
then $q\bar{q}$ may not form bound states. This is the conventional
mechanism of quarkonia disintegration. Due to this melting, the yield
of quarkonia will be \textit{suppressed}. This was proposed as a
signature of QGP and has been observed experimentally
\cite{Matsui:1986dk,Abreu:2000ni}. However, there are other factors 
too that can lead to the suppression of $J/\psi$ because of which it 
has not been possible to use $J/\psi$ suppression as a clean signal for QGP. 

In this paper, we propose a novel mechanism of quarkonia disintegration 
via QCD $Z(3)$ domain walls. These walls appear as 
topological defects due to spontaneous breaking  of $Z(3)$ symmetry in 
QGP \cite{Bhattacharya:1992qb,West:1996ej,Boorstein:1994rc}. The thermal
expectation value of Wilson loop (Polyakov loop) acts as the order
parameter for confinement-deconfinement phase transition taking zero
value in the confining phase (corresponding to infinite free energy of
a test quark) and a non-zero value in the QGP phase (with finite
free energy of a test quark). Polyakov loop transforms non-trivially
under $Z(3)$ transformations, hence its non-zero expectation value
leads to spontaneous breaking of $Z(3)$ symmetry in the QGP 
phase \cite{Polyakov:1978vu,McLerran:1981pb}. With the possibility of
realization of the QGP phase in RHIC and LHC experiments, we have the
real opportunity to study topological domain walls, resulting from
this spontaneous $Z(3)$ symmetry breaking, in laboratory.  The formation 
and evolution of these walls have been discussed in context of RHIC 
experiments \cite{Gupta:2010pp,Gupta:2011ag}. The associated {\it QGP} 
string formation \cite{Layek:2005fn} has also been discussed by some of us. 
It is important to mention here that such topological defects
invariably form during a phase transition. The formation process of
topological defects is governed by formation of a sort of domain structure
during a phase transition, and is usually known as the {\it Kibble mechanism}
\cite{kbl}. The network of defects formed depends on the details of phase 
transition only through the correlation length. In fact defect distribution,
e.g. defect density, per correlation volume is universal and depends
only on the symmetry breaking pattern and space dimensions.  

 Questions have been raised on the \textit{reality} of these $Z(3)$ 
domains. The existence of these Z(3) vacua
becomes especially a non-trivial issue when considering the presence
of dynamical quarks. The effect of quarks on $Z(3)$ symmetry and $Z(3)$
interfaces etc. has been discussed in detail in the literature 
\cite{qurk1,qurk2}. It has also been argued that the $Z(3)$ symmetry 
becomes meaningless in the presence of quarks \cite{qurk1}. Other 
view-point, as advocated in many papers, asserts that one can take the 
effect of  quarks in terms of explicit breaking of $Z(3)$ symmetry 
\cite{qurk2,psrsk,psrsk2}. We follow this approach and assume that the 
effects of dynamical quarks can be incorporated by introducing explicit 
symmetry breaking terms in the effective potential for the Polyakov loop. 
This makes Z(3) domain wall dynamical with pressure difference between 
the two different vacua being non-zero. This will lead to asymmetric 
profile of the Polyakov loop. In the present paper, we will ignore these 
{\it asymmetry} effects due to dynamical quarks, and will continue using
Z(3) interfaces without any explicit symmetry breaking term. In a future
work we will come back to include the effects of explicit symmetry breaking.
 
 We mention recent lattice results in 
ref.(\cite{Deka:2010bc}) which indicate strong possibility of the 
existence of these Z(3) domains at high temperatures in the presence 
of dynamical quarks. These results suggest that (metastable)
Z(3) domains appear at temperatures above about 700 MeV. Though, we
stress that it does not look appropriate to take these results as 
conclusive, especially the quantitative part. Thus one would like to 
consider the possibility that Z(3) vacua may persist for somewhat 
lower temperatures also as discussed in this paper. In any
case, the mechanism discussed here provides additional source
of disintegration for $J/\psi$ even at high temperature. It is important
to note that our mechanism will lead to disintegration of
Upsilon also which will be relevant even at 700 MeV. We will present
this study in a future work. 

In case of early universe, these $Z(3)$ walls can lead to baryon 
inhomogeneity generation \cite{Layek:2005zu}. It was shown in 
ref.\cite{KorthalsAltes:1994if} that background gauge field $A_0$
associated with \textit{generalized} $Z(N)$ interfaces can lead to 
spontaneous CP violation in SM, MSSM and SUSY models, which, in turn, 
can lead to baryogenesis in the early universe. A detailed quantitative 
analysis of this spontaneous CP violation was done in 
\cite{Atreya:2011wn}, in the context of quark/antiquark scattering 
from $Z(3)$ walls in the QGP phase. The main approach
followed in refs.\cite{KorthalsAltes:1994if,Atreya:2011wn} was 
based on the assumption that the profile of the Polyakov loop
order parameter $l(x)$ corresponds to a sort of condensate of the background 
gauge field $A_0$ (in accordance with the definition of the Polyakov loop). 
This profile of the background gauge field can be calculated from the profile 
of $l(x)$. Such a gauge field configuration in the Dirac equation
leads to different potentials for quarks and antiquarks, leading
to spontaneous CP violation in the interaction of quarks and antiquarks from
the Z(3) wall. This spontaneous CP violation was first 
discussed by Altes et al. \cite{KorthalsAltes:1994be,KorthalsAltes:1994if} 
in the context of the universe and in ref. \cite{KorthalsAltes:1992us} for 
the case of QCD. In \cite{Atreya:2011wn}, the profile of Polyakov loop 
$l(x)$ between different Z(3) vacua was used (which was obtained by using 
specific effective potential for $l(x)$ as discussed in
\cite{psrsk,psrsk2}) to obtain the profile of $A_0$. 
This background $A_0$ configuration acts as a potential for
quarks and antiquarks. It was shown in ref.\cite{Atreya:2011wn} that 
the quarks have significantly different reflection coefficients than 
anti-quarks and the effect is stronger for heavier quarks. 
For a discussion of calculation of $A_0$ profile,
see ref. \cite{Atreya:2011wn}. 
 
In this paper, we discuss the effect of this spontaneous CP violation on 
the propagation of quarkonia in the QGP medium, in particular, the 
$J/\psi$ meson. $J/\psi$ are produced in the initial stages of 
relativistic heavy ion collisions. As these are heavy  mesons 
$(m\sim 3 GeV)$, they are never in equilibrium with the QGP medium 
produced in present heavy-ion collision experiments. However, there are 
finite $T$ effects (like Debye screening etc.) affecting its motion in a 
thermal bath. We ignore these effects initially and comment on it towards the 
end. Note that if the Debye length is larger, then the conventional mechanism 
of $J/\psi$ melting does not work. As we will argue, for large Debye 
screening, our mechanism of $J/\psi$ disintegration works better as any 
possible screening of the domain wall over the relevant length scale of 
$J/\psi$ will be small. If a domain wall is present in the QGP, then a 
$J/\psi$ moving through the wall will have a non-trivial interaction 
with it. Due to the CP violating effect of the interface on quark scattering, 
$c$ and $\bar{c}$ in $J/\psi$ experience different color forces depending on 
the color of the quark and the color composition of the wall. This not only 
changes the color composition of $c\bar{c}$ bound state (from color
singlet to color octet state) but also facilitates its transition to 
higher excited states (for example $\chi$ states). Color octet quarkonium
states are unbound (also, the $\chi$ state has larger size than $J/\psi$ and
the Debye length), hence they will dissociate in the QGP medium. This
summarizes the basic physics of our model discussed in this paper for 
quarkonia disintegration due to $Z(3)$ walls. 

 The paper is organized in the following manner. In section II we discuss
 the interaction of $J/\psi$ with the background gauge field $A_0$ arising
 from the profile of $l(x)$ and discuss its color excitations. Subsequently 
 we consider spatial excitations of $J/\psi$ and calculate the 
 disintegration probability. Section 
 III discusses results, and conclusions are presented in Sect.IV.

\section{Interaction of $J/{\psi}$ with a Z(3) wall}

 In our model, $J/\psi$ interacts with the gauge field $A_0$ corresponding
 to the $l(x)$ profile of the Z(3) wall. This allows for the possibility
 of color excitations of $J/\psi$ as well as the spatial excitations of
 its wave function. First we discuss the possibility of  color 
 excitations of $J/\psi$. Subsequently, we will discuss spatial excitations
 of $J/\psi$.

\subsection{Color excitation of $J/{\psi}$}

We work in the rest frame of $J/\psi$ and consider the domain wall coming 
and hitting the $J/\psi$ with a velocity $v$ along $z$-axis. The gauge 
potential and coordinates are appropriately Lorentz transformed as

\begin{subequations}\label{eq:lrntz}
\begin{align}
A_{0}(z)\rightarrow A_{0}^{\prime}(z^{\prime}) &= \gamma\left(A_{0}(z)-vA_{3}(z)
\right) \label{eq:a0trnsfrm}\\
A_{3}(z)\rightarrow A_{3}^{\prime}(z^{\prime}) &= \gamma\left(A_{3}(z)-vA_{0}(z)
\right) \label{eq:aitrnsfrm}\\
z &= \gamma\left(z^{\prime}+vt^{\prime}\right).
\end{align}
\end{subequations}

We assume that there is no background vector potential, $A_{i}(z) = 0~; 
i=1,2,3$. $A_{3}^{\prime}$ obtained from Eqn.. (\ref{eq:aitrnsfrm})
has only $z^{\prime}$ dependence, so it does not produce any color magnetic
field. Further, using the non-relativistic approximation of the Dirac
equation one can see that the perturbation terms in the Hamiltonian 
(say, $H^1(A_3^{\prime})$) involving $A_3^{\prime}$ are suppressed compared 
to the perturbation term ($H^1(A_0^{\prime})$) involving $A_0^{\prime}$
at least by a factor

\begin{equation}\label{eq:a3}
{H^1(A_3^{\prime}) \over H^1(A_0^{\prime})} \sim 
{v \over c} {1 \over m_c r_{J/\psi}}
\end{equation}

where $r_{J/\psi}$ is the size of the $J/\psi$ wave function and $m_c$ is
the charm quark mass. As we will see, the largest value of $v/c$ we consider
is 0.20 - 0.24 (above which transition amplitude becomes too large to
trust first order perturbation approximation). With $r_{J/\psi} \simeq
0.4$ fm, the suppression factor in Eqn.(2) is of order 10 \%. Thus we neglect
perturbation due to $A_3^{\prime}$ and only consider perturbation due to
$A_0$ as given by Eqn.(1a). We use first order time
dependent perturbation theory to study the excitation
of $J/\psi$ due to the background $A_{0}$ profile and consider the transition
of $J/\psi$ from initial energy eigenstate $\psi_{i}$ with energy $E_{i}$ to
the final state $\psi_{j}$ with energy $E_{j}$. The transition amplitude is
given by

\begin{equation}
\mathcal{A}_{ij} = \delta_{ij} - 
~i\int_{t_{i}}^{t_{f}}\langle\psi_{j}|\mathcal{H}_{int}|\psi_{i}\rangle 
e^{i(E_{j}-E_{i})t}dt.
\label{eq:amp}
\end{equation}

We take incoming quarkonia to be a color singlet state. The 
interaction of the quarkonia with the wall is written as

\begin{subequations}\label{eq:hint}
\begin{align}
\mathcal{H}_{int} &= V^{q}(z_{1}^{\prime})\otimes\mathbb{1}^{\bar{q}}  + \mathbb{1}^{q}\otimes V^{\bar{q}}(z_{2}^{\prime})\\
\text{with}~~ V^{q,\bar{q}}(z_{1,2}^{\prime}) &= gA_{0}^{\prime q,\bar{q}}(z_{1,2}^{\prime}),
\end{align}
\end{subequations}
where $A_{0}^{\prime q,\bar{q}}(z_{1,2}^{\prime})$ is the background field 
configuration in the rest frame of $J/\psi$. $z_{1}^{\prime}$ and $z_{2}^{\prime}$ 
are the coordinates of $q$ and $\bar{q}$ in quarkonia and $g$ is the gauge 
coupling. The gauge potential $A_{0}$ is taken in the diagonal gauge as
\begin{equation}
A_{0} = \frac{2\pi T}{g}\left(a\lambda_{3} + b\lambda_{8}\right),
\label{eq:a0}
\end{equation}
where $\lambda_{3}$ and $\lambda_{8}$ are the Gell-Mann matrices. Under CP, 
$A_{0}\rightarrow -A_{0}$, hence $A_{0}^{\bar{q}} = -A_{0}^{q}$. Now, both the 
initial and the final states have a spatial, spin and color part. The incoming 
quarkonia is a color singlet while outgoing state could be a singlet or an 
octet. Using Eqn. (\ref{eq:hint}), (\ref{eq:a0}) and extracting only the color 
part of interaction, we get
\begin{equation}\label{eq:transamp}
\begin{split}
\langle\psi_{out}|\mathcal{H}_{int}|\psi_{singlet}\rangle &= 
\langle\psi_{out}|gA_{0}^{\prime q}(z_{1}^{\prime})\otimes \mathbb{1}^{\bar{q}}|\psi_{singlet}\rangle \\
& + \langle\psi_{out}|\mathbb{1}^{q}\otimes gA_{0}^{\prime\bar{q}}(z_{2}^{\prime})|\psi_{singlet}\rangle.
\end{split}
\end{equation}
\hspace*{0.5cm}The color singlet state of $J/\psi$ is written as
\begin{equation}
\begin{split}
|\psi_{singlet}\rangle &= 
\frac{1}{\sqrt{3}}\Biggl[\begin{pmatrix}
 1 \\
 0 \\
 0 \\
 \end{pmatrix}^{q}\otimes \begin{pmatrix}
                           1 \\
                           0 \\
                           0 \\
                           \end{pmatrix}^{\bar{q}} +
\begin{pmatrix}
 0 \\
 1 \\
 0 \\
 \end{pmatrix}^{q}\otimes \begin{pmatrix}
                           0 \\
                           1 \\
                           0 \\
                           \end{pmatrix}^{\bar{q}} \\
& \quad \quad  \quad \quad \quad + \begin{pmatrix}
                                    0 \\
                                    0 \\
                                    1 \\
                                  \end{pmatrix}^{q}\otimes \begin{pmatrix}
                                                            0 \\
                                                            0 \\
                                                            1 \\
                                                          \end{pmatrix}^{\bar{q}}\Biggr].
\label{eq:clmvctr}
\end{split}
\end{equation}
\hspace*{0.5cm} If the outgoing state is also a singlet then, each term on RHS 
of Eqn. (\ref{eq:transamp}) is zero due to the traceless nature of $A_{0}$. 
Eqn. (\ref{eq:amp}) gives $\mathcal{A}_{ij} = 1$ for ground state $(i=j)$. 
(Meaning, one will then need to resort to 2nd order perturbation theory
for consistency). For 
higher orbital states $(i\neq j)$, amplitude is identically zero. A color 
octet state like $|r\bar{g}\rangle$, can be written as
\begin{equation}
|r\bar{g}\rangle = \begin{pmatrix}
                    1 \\
                    0 \\
                    0 \\
                   \end{pmatrix}^{q}\otimes \begin{pmatrix}
                                          0 \\
                                          0 \\
                                          1 \\
                                         \end{pmatrix}^{\bar{q}}.
\label{eq:rgstate}
\end{equation}
\hspace*{0.5cm} For such an outgoing state each term on RHS of Eqn. 
(\ref{eq:transamp}) again vanishes identically because of the diagonal form of 
$A_{0}$, resulting in zero transition probability. Same argument leads to zero 
transition probability to all other octet states with similar color content, 
viz. $b\bar{g},~b\bar{r},~g\bar{r},~g\bar{b},~r\bar{b}$. There are only two 
states with non-zero color contribution to transition probability. They are  
\begin{equation}
\begin{split}
|r\bar{r}-b\bar{b}\rangle &= 
\frac{1}{\sqrt{2}}\Biggl[\begin{pmatrix}
 1 \\
 0 \\
 0 \\
\end{pmatrix}^{q}\otimes \begin{pmatrix}
                          1 \\
                          0 \\
                          0 \\
                        \end{pmatrix}^{\bar{q}} 
- \begin{pmatrix}
   0 \\
   1 \\
   0 \\
  \end{pmatrix}^{q}\otimes \begin{pmatrix}
                            0 \\
                            1 \\
                            0 \\
                          \end{pmatrix}^{\bar{q}}\Biggr]
\label{eq:rbstate}
\end{split}
\end{equation}
and
\begin{equation}
\begin{split}
|r\bar{r}+b\bar{b}-2g\bar{g}\rangle &= 
\frac{1}{\sqrt{6}}\Biggl[\begin{pmatrix}
 1 \\
 0 \\
 0 \\
\end{pmatrix}^{q}\otimes \begin{pmatrix}
                          1 \\
                          0 \\
                          0 \\
                        \end{pmatrix}^{\bar{q}} 
+ \begin{pmatrix}
   0 \\
   1 \\
   0 \\
  \end{pmatrix}^{q}\otimes \begin{pmatrix}
                            0 \\
                            1 \\
                            0 \\
                          \end{pmatrix}^{\bar{q}}\\
& \quad \quad \quad \quad - 2\begin{pmatrix}
                              0 \\
                              0 \\
                              1 \\
                            \end{pmatrix}^{q}\otimes \begin{pmatrix}
                                                      0 \\
                                                      0 \\
                                                      1 \\
                                                    \end{pmatrix}^{\bar{q}}\Biggr].
\label{eq:rbgstate}
\end{split}
\end{equation}
\hspace*{0.5cm}Using Eqn. (\ref{eq:rbstate}) and (\ref{eq:rbgstate}) in 
conjunction with Eqn. (\ref{eq:a0}),(\ref{eq:lrntz}) and (\ref{eq:transamp}), 
we get the color part of transition probability as
\begin{subequations}\label{eq:octetamp}
\begin{align}
\langle r\bar{r}-b\bar{b}|\mathcal{H}_{int}|\psi_{singlet}\rangle = {1 \over
\sqrt{6}} (A_{0}^{r} - A_{0}^{b}) \label{eq:rbamp}\quad \text{and} \quad \\
\langle r\bar{r}+b\bar{b}-2g\bar{g}|\mathcal{H}_{int}|\psi_{singlet}\rangle 
= {1 \over \sqrt{18}} (A_{0}^{r} + A_{0}^{b} - 2A_{0}^{g}),\label{eq:rbgamp},
\end{align}
\end{subequations}

where, $A_{0}^{r},~A_{0}^{b} ~\text{and}~ A_{0}^{g}$ are the diagonal 
components of the matrix $A_{0}^{\prime}\left(z_{1}^{\prime}\right)-
A_{0}^{\prime}\left(z_{2}^{\prime}\right)$. Eqn. (\ref{eq:rbamp}) and 
(\ref{eq:rbgamp}) are the effective interactions that lead to the 
excitations of incoming $J/\psi$ (in the color singlet state of $c\bar{c}$) 
to the corresponding octet state. Due to repulsive Coulombic interaction
of $q$ and $\bar{q}$ in the octet representation, one may expect that
$J/\psi$ may disintegrate while traversing through a $Z(3)$ wall purely
by color excitation. However, we will see in the next section that this
is not so and one needs to also consider spatial excitation of $J/psi$ 
due to $Z(3)$ wall.

\subsection{Spatial excitations of $J/\psi$}

We now consider the spatial excitations. The spatial part of the states is 
decided by the potential between $c\bar{c}$ in $J/\psi$ which is taken as,

\begin{equation}\label{eq:potnl}
V\left(|\vec{r}_{1}-\vec{r}_{2}|\right) = 
-\frac{\alpha_{s}C_F }{|\vec{r}_{1}-\vec{r}_{2}|} + 
C_{cnf} ~\sigma |\vec{r}_{1}-\vec{r}_{2}|
\end{equation}

where $\alpha_{s}$ is the strong coupling constant and $\sigma$ is the string 
tension. For $J/\psi$, we will use charm quark mass $m_{c} = 1.28 $ GeV,
$\alpha_{s} = \pi/12$, and $\sigma = 0.16~GeV^{2}$ 
\cite{satz,Giannuzzi:2009gb}. $C_F$ is the color factor depending on the 
representation of the $c\bar{c}$ state. $C_F = 4/3$ for singlet state, 
while $C_F = -1/6$ for the octet states showing the repulsive nature of the 
Coulombic part of the interaction for the octet states. $C_{cnf}$ denotes 
the representation dependence of the confining part of the potential. For
general sources, this factor follows Casimir scaling \cite{csmrscl} for
the string tension. For $J/\psi$ in singlet representation, $C_{cnf} = 1$ 
with the value of $\sigma$ used here \cite{satz,Giannuzzi:2009gb}. 
It is not clear what should be the
value of $C_{cnf}$ if  $c\bar{c}$ are in the octet representation.
As the Coulombic part of the potential is repulsive for the octet
state of $c\bar{c}$ (with $C_F = -1/6$), it is not clear if there should be 
a confining part of the potential at all in this case for large distances.
Early lattice simulations had indicated some possibility of mildly rising
potential for the confining part for $q\bar{q}$ in octet 
representation \cite{octerearly}. However, recent simulations do not show 
any such possibility. At large distances, the net potential between a 
$q$ and $\bar{q}$ in color octet state appears to be
independent of distance \cite{octetnew}. With the repulsive Coulombic
part, this implies a very small value for $C_{cnf}$ for the confining
part. For our purpose it suffices to assume that in the octet representation,
$J/\psi$ becomes unbound, having repulsive interaction at short distances. 

  We have seen above that the form of $A_0$ in Eqn.(5) only allows for 
transition from color singlet to two of the color octet states given in 
Eqns.(9),(10). As we discussed above, $c\bar{c}$ in color octet state is
unbound. Thus our task should be to consider transition from initial 
color singlet $J/\psi$ to unbound state of $c\bar{c}$, say in plane waves. 
However, this also does not look correct as the initial $J/\psi$ (in the 
color singlet state) transforms to a color octet state only as it traverses 
the $Z(3)$ wall (as coefficients $a$ and $b$ in Eqn.(5) undergo spatial 
variations). Thus during the early part of the passage of $J/\psi$ through 
the wall, it should be dominantly in the singlet state (which is a bound 
state) and it will be incorrect to consider transition to unbound, plane
wave states of $c\bar{c}$ at this stage. Only at later stages, when the
octet component is dominant, it may be appropriate to consider repulsive
potential in Eqn.(12), and unbound $c\bar{c}$ states for the transition
probability. This means that the perturbation term should appropriately 
account for the growth of octet component for the potential in Eqn.(12), 
along with a continuing singlet component with corresponding singlet 
potential in Eqn.(12).  This clearly is a complex issue, and a proper 
account of appropriate potential for this type of evolution of $J/\psi$ 
cannot be carried out in simple approximation scheme considered here.
We make a simplifying assumption that $J/\psi$ becomes unbound only when it 
transforms to the octet representation {\it after} its interaction with 
the $Z(3)$ wall. Until then it is assumed to be in the
color singlet representation. Thus, in the calculations of the
spatial excitation of the $J/\psi$ state below, we use the
$c\bar{c}$ potential (Eqn.(12)) in the color singlet representation.
The underlying physics is that incoming $J/\psi$ is in the color
singlet state, it interacts with Z(3) wall which excites it to
higher state (spatial excitation), still in color singlet potential.
While traversing the Z(3) wall, and undergoing this spatial
excitation, the $J/\psi$ state also transforms to color octet state.
The final state, after traversing the Z(3) wall, is spatially
excited state in color octet representation, and our calculations
give probability for this final state. This final octet state is unbound 
and hence such excited $J/\psi$ disintegrates. We emphasize that at this 
stage, our aim is to point out the new possibilities of 
disintegration of $J/\psi$ with Z(3) walls and this simplifying assumption
should not affect our qualitative considerations and approximate 
estimates. We hope to give a more complete treatment in future.
Thus, we continue to use the color singlet potential in Eqn.(12), while
considering the spatial excitation of $J/\psi$. 

Since the potential is central, we perform coordinate transformations
\begin{align}\label{eq:cordtrns}
\vec{R}_{cm} &= \frac{\vec{r}_{1}+\vec{r}_{2}}{2} & 
\text{and}\quad \quad \vec{r} &= \vec{r}_{1}-\vec{r}_{2},
\end{align}

where, $ \vec{r}$ is the relative coordinate between $q$ and $\bar{q}$. 
$\vec{R}_{cm}$ is the center of mass of $J/\psi$. Using Eqn. 
(\ref{eq:cordtrns}) with Eqn. (\ref{eq:lrntz}), we get

\begin{equation}\label{eq:radiala0}
A_{0}^{r} = \gamma A_{0}^{11} \bigl[\gamma(z^{\prime}_{1}+vt^{\prime})\bigr] 
- \gamma A_{0}^{11} \bigl[\gamma(-z^{\prime}_{2}+vt^{\prime})\bigr].
\end{equation}

$z^{\prime}_{1}$ and $z^{\prime}_{2}$ are written in terms of $\vec{R}_{cm}$ 
and $\vec{r}$. Similar expressions can be obtained for $A_{0}^{b}$ and 
$A_{0}^{g}$. In the above coordinates, the $J/\psi$ wave function 
is $\Psi(\vec{R}_{cm})\psi(\vec{r})$. For simplicity, we assume that the center 
of mass motion remains unaffected by the external perturbation. Then 
$\Psi(\vec{R}_{cm})$ has the plain wave solution, while $\psi(\vec{r})$ can be 
written $\psi(r,\theta,\phi) = \psi(r)Y_{l}^{m}\left(\cos\theta,\phi\right)$. 
As $J/\psi$ is the $l = 0$ state, we have

%
\begin{align}\label{eq:wvfnctn}
\psi_{i} &= \psi(r)Y_{0}^{0} & \text{and} \quad \psi_{j} &= \psi_{n}\left(r\right)Y_{l}^{m}\left(\cos\theta,\phi\right).
\end{align}
%

The radial part, $\psi(r)$, is obtained by solving radial part 
of Schr$\ddot{o}$dinger equation with effective potential given by
\begin{equation}\label{eq:effpotnl}
V\left(r\right) = -\frac{\alpha_{s} C_F}{r} + 
C_{cnf}~ \sigma r + \frac{l(l+1)}{2\mu r^{2}}
\end{equation}
where $\mu$ is the reduced mass. When we use Eqn. (\ref{eq:octetamp}), 
(\ref{eq:radiala0}) and (\ref{eq:wvfnctn}) in Eqn. (\ref{eq:amp}), we get one 
of the terms as
\begin{equation}\label{eq:radial}
\begin{split}
\int_{-\infty}^{\infty}\psi_{j}^{*}A_{0}^{r}\psi_{i}~d\vec{r}_{1}d\vec{r}_{2} 
&= \int_{0}^{\infty}\int_{-1}^{1}\int_{0}^{2\pi}\psi^{*}_{n}(r)
Y_{l}^{m*} \left(\cos\theta,\phi\right)A_{0}^{r}\\ 
& \quad \quad \quad \quad Y_{0}^{0}\psi_{100}(r)~r^{2}~dr d(\cos\theta) d\phi.
\end{split}
\end{equation}

In the above equation, we have ignored the motion of the center of 
mass of charmonium and have considered only the relative coordinate. Under 
$\cos\theta \rightarrow -\cos\theta, A_{0}^{r} \rightarrow -A_{0}^{r}$ and 
$\psi_{i}$ does not change. So if $Y_{l}^{m}\left(\cos\theta,\phi\right) = 
Y_{l}^{m}\left(-\cos\theta,\phi\right)$ then RHS of Eqn. (\ref{eq:radial}) is 
zero. Thus we do not get any transition to a state which is symmetric under 
$\cos\theta \rightarrow -\cos\theta$. This has very important significance. 
While the color part prohibits the transition to singlet final states, the 
space dependence of interaction forbids the transition to the 
$l = 0$ state (in color octet). Thus we see that purely color excitation of 
$J/\psi$ due to $A_0$ field of a domain wall is not possible. The excitation 
is possible to the first excited state of an octet (like an `octet $\chi$' 
state). As the excited state will have a radius larger than the $l = 0$ 
state it is more prone to melting in the medium, (though with color octet
composition, the final state becomes unbound anyway).

\begin{figure}[!htp]
\begin{center}
\includegraphics[width=0.45\textwidth]{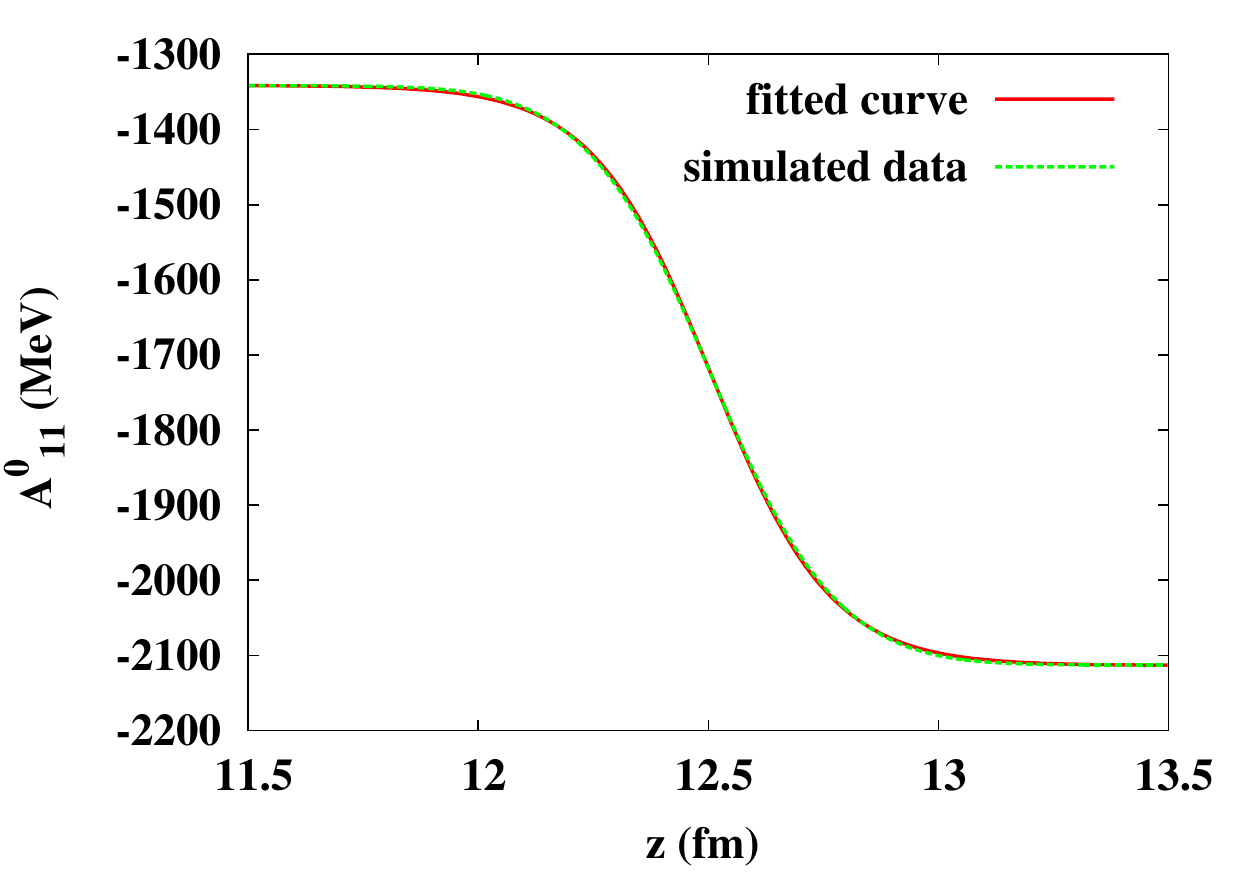}
\caption{(Color online) $A_{0}$ profile across the Z(3) domain wall 
for $T = $ 400 MeV. Only $(1,1)$ component is shown. Other components 
are similar. See ref. \cite{Atreya:2011wn} for details.}
\label{fig:a0}
\end{center}
\end{figure}

\section{Results}

We numerically compute the integral given in Eqn. 
(\ref{eq:amp}) with various parameters given after Eqn.(12).
The profile of $A_{0}$ is calculated from the
profile of the Polyakov loop order parameter for a Z(3) domain wall
at a temperature $T = $400 MeV (as a sample value). The details of this 
are given in ref. \cite{Atreya:2011wn}. As explained there, the resulting 
profile is very well fitted by the functional form $p\tanh(qx+r)+s$, 
see Fig. \ref{fig:a0}. 

\begin{figure}[!htp]
\begin{center}
\includegraphics[width=0.45\textwidth]{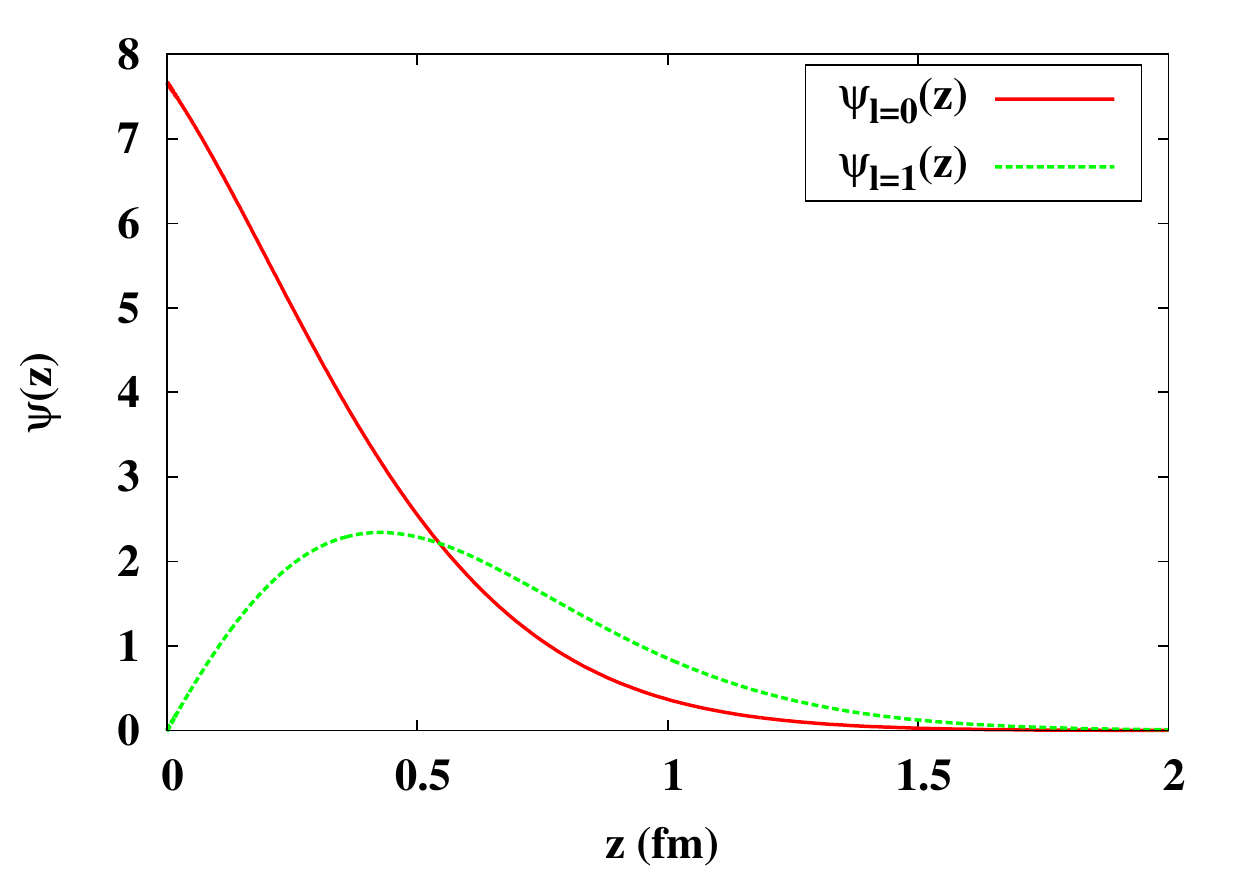}
\caption{(Color online) Wave functions for $J/\psi~(l=0)$ and 
$\chi~(l=1)$ states.}
\label{fig:wvfnctn}
\end{center}
\end{figure}

We calculated the wave functions for various states of $c\bar{c}$ with
the complete potential given by Eqn. (\ref{eq:effpotnl}). 
For the calculation of the wave-functions for 
various states of $c\bar{c}$ we have used Numerov method for solving
the Schr$\ddot{o}$dinger equation. We have also used energy 
minimization technique to get the wave functions and the bound state 
energy and the results obtained by both the methods match very well.
Fig. (\ref{fig:wvfnctn}) shows the radial part of the
wave function for the $l=0,1$ states of charmonium. The bound state 
contributions to the energy (excluding the rest mass of quarks) are 
found to be $E_{0} =0.447$ GeV for $J/\psi$ and $E_{0} = 0.803 $ GeV 
for $\chi$ state ($l = 1$). We see from Fig.2 that 
the radius of $J/\psi$ is about $0.5~fm$ while that for $\chi$ is about 
$0.8~fm$. Debye length in QGP  at $T = 200$ MeV is $r_{d} \sim 0.6$ fm 
and smaller at higher temperatures. Thus $\chi$ state is unstable and 
it should melt easily in the medium (apart from the fact that in color octet
state it also becomes unbound). Fig.3 shows the 
combined probability of transition to both the color octet $\chi$ states 
(Eqns.(9),(10)) for an 
incoming $J/\psi$ with different velocities moving normal to the 
domain wall. As we see, the probability rapidly rises as a function
of velocity. However, for large velocities the probability of
transition becomes large making first order perturbation approximation
insufficient, and one needs more reliable estimates. 
Thus, the plot in Fig.3 should be trusted only for
small velocities. Nonetheless, the trend at higher velocities
strongly suggests that most of $J/\psi$ will disintegrate while
interacting with $Z(3)$ walls.

\begin{figure}[!htp]
\begin{center}
\includegraphics[width=0.45\textwidth]{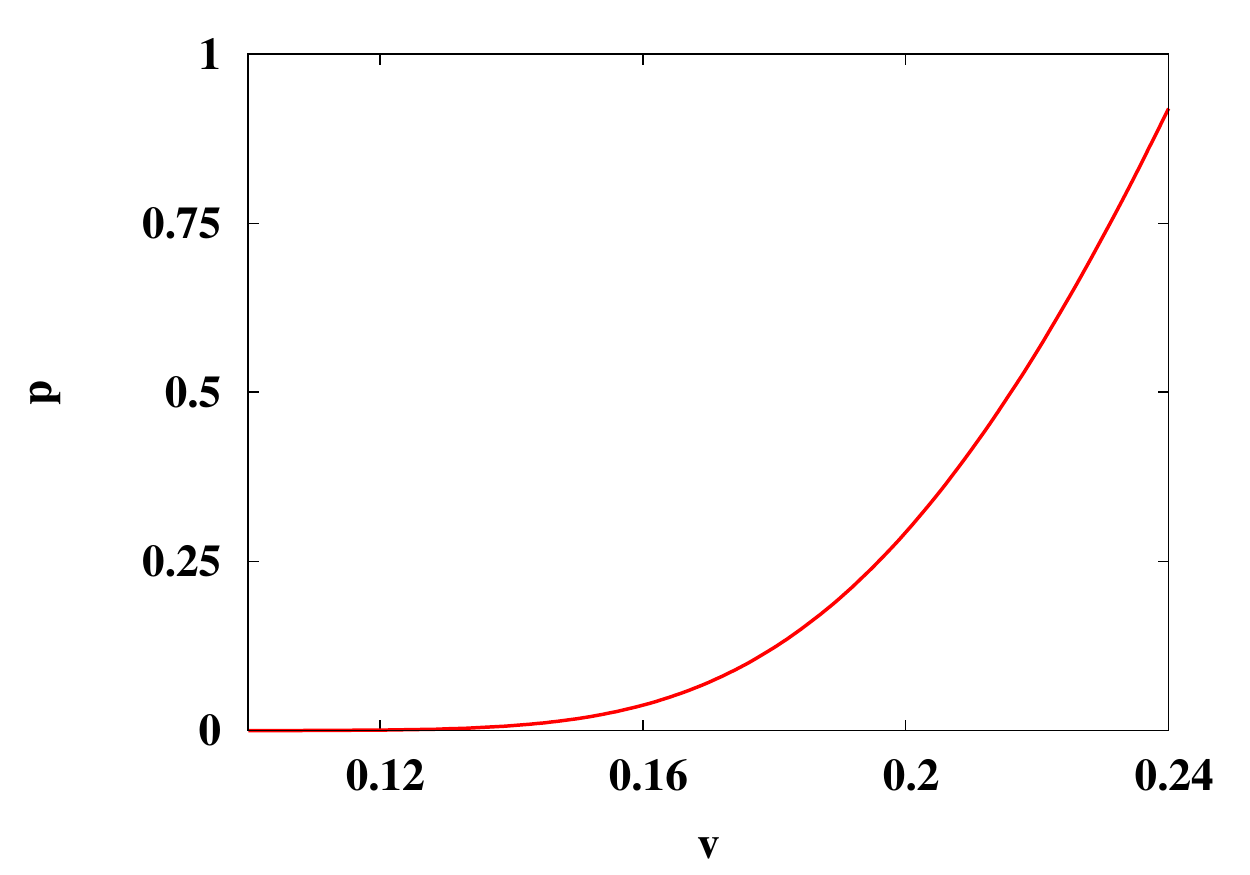}
\caption{(Color online) Probability $p$ of transition of $J/\psi$ 
to color octet $\chi$ states vs. its velocity $v$. Note that the 
probability rapidly rises with $v$.}
\label{fig:prob}
\end{center}
\end{figure}

\section{Conclusions}

These results  show that on interaction with a $Z(3)$ domain 
wall, a $J/\psi$ particle will make an excitation to a higher orbital 
state in color octet representation which is unbound and will readily melt 
in the surrounding QGP medium. At higher energies, the transition probability  
keeps increasing, making the first order perturbation theory inapplicable  
and the results are not trustworthy. Nonetheless, this implies that 
at higher energies, almost all $J/\psi$ are expected to disintegrate
in this manner. This strong $P_T$ dependence of $J/\psi$ disintegration 
probability is a distinctive signature of our model wherein the probability 
of disintegration of $J/\psi$ is enhanced with higher $P_T$. This can be
used to distinguish this mechanism from the conventional Debey screening
suppression. A very crucial point in the entire discussion is the Debye 
screening of the $A_0$ profile of the domain wall itself as it carries
color. At temperature 400 MeV, the domain wall has a thickness of $\sim 
1.5~fm$ and the Debye radius for QGP is $\sim 0.7~fm$. This means that 
Debye screening will be effective outside a sphere of diameter $\sim 1.5~fm$. 
So we do not expect the domain wall to be significantly Debye screened. 
In the above discussion, we have completely ignored the effects of a thermal 
bath (QGP medium) on the potential (Eqn. \ref{eq:potnl}) between $c\bar{c}$ 
(\cite{satz,Digal:2005ht}). However as these effects make the potential between 
$c\bar{c}$ weaker, the charmonium state \textit{swells}. So it will be even 
easier for the interaction to break these bound states. These temperature 
effects will also be crucial for other heavier $q\bar{q}$ states like 
bottomonium as they have large binding energies. Another important aspect 
which has been ignored for the sake of simplicity, in the above calculations, 
is the question of the center of mass motion. This assumption is correct only 
in an average sense as the average force $(\Delta V/\Delta z)$ acting on $c$ 
and $\bar{c}$ vanishes. This averaging is done over the thickness $\Delta z$, 
which is the thickness of the domain wall itself. However as the instantaneous 
force $\left(\partial V/\partial z\right)$ is non-zero, there is a non-zero 
instantaneous acceleration of the center of mass. A more detailed analysis of 
the problem is required to incorporate all these details. One also needs to
include the effects of dynamical quarks leading to explicit breaking of
Z(3) symmetry. We mention that such a disintegration of $J/\psi$ from a color 
electric field may not necessarily come from a background domain wall arising 
in QGP medium. In a thermal medium there are always statistical fluctuations. 
These gluonic fluctuations will have energy of order $\sim T$. Depending on 
the correlation length of the fluctuation, a $J/\psi$ passing through it may 
disintegrate via the mechanism discussed above. It would be interesting to 
study the effect of these thermal gluonic fluctuations on the spectrum of 
mesons.

\section*{Acknowledgment}
We are very thankful to Arpan Das for useful discussions and for 
pointing out an error in calculations. We thank Sanatan Digal, 
Rajarshi Ray, Ranjita Mohapatra, Saumia P.S., Souvik Banarjee, and 
Ananta P. Mishra for useful discussions. We thank an anonymous referee 
for pointing out the Casimir scaling factor and the Numerov method, and 
for pointing out an important error in Casimir factors.


\begin{thebibliography}{18}
\expandafter\ifx\csname natexlab\endcsname\relax\def\natexlab#1{#1}\fi
\expandafter\ifx\csname bibnamefont\endcsname\relax
  \def\bibnamefont#1{#1}\fi
\expandafter\ifx\csname bibfnamefont\endcsname\relax
  \def\bibfnamefont#1{#1}\fi
\expandafter\ifx\csname citenamefont\endcsname\relax
  \def\citenamefont#1{#1}\fi
\expandafter\ifx\csname url\endcsname\relax
  \def\url#1{\texttt{#1}}\fi
\expandafter\ifx\csname urlprefix\endcsname\relax\def\urlprefix{URL }\fi
\providecommand{\bibinfo}[2]{#2}
\providecommand{\eprint}[2][]{\url{#2}}

\bibitem[{\citenamefont{Matsui and Satz}(1986)}]{Matsui:1986dk}
\bibinfo{author}{\bibfnamefont{T.}~\bibnamefont{Matsui}} \bibnamefont{and}
  \bibinfo{author}{\bibfnamefont{H.}~\bibnamefont{Satz}},
  \bibinfo{journal}{Phys.Lett.} \textbf{\bibinfo{volume}{B178}},
  \bibinfo{pages}{416} (\bibinfo{year}{1986}).

\bibitem[{\citenamefont{Abreu et~al.}(2000)}]{Abreu:2000ni}
\bibinfo{author}{\bibfnamefont{M.}~\bibnamefont{Abreu}} \bibnamefont{et~al.}
  (\bibinfo{collaboration}{NA50 Collaboration}), \bibinfo{journal}{Phys.Lett.}
  \textbf{\bibinfo{volume}{B477}}, \bibinfo{pages}{28} (\bibinfo{year}{2000}).

\bibitem[{\citenamefont{Bhattacharya et~al.}(1992)\citenamefont{Bhattacharya,
  Gocksch, Korthals~Altes, and Pisarski}}]{Bhattacharya:1992qb}
\bibinfo{author}{\bibfnamefont{T.}~\bibnamefont{Bhattacharya}},
  \bibinfo{author}{\bibfnamefont{A.}~\bibnamefont{Gocksch}},
  \bibinfo{author}{\bibfnamefont{C.}~\bibnamefont{Korthals~Altes}},
  \bibnamefont{and} \bibinfo{author}{\bibfnamefont{R.~D.}
  \bibnamefont{Pisarski}}, \bibinfo{journal}{Nucl. Phys.}
  \textbf{\bibinfo{volume}{B383}}, \bibinfo{pages}{497} (\bibinfo{year}{1992})
 

\bibitem[{\citenamefont{West and Wheater}(1997)}]{West:1996ej}
\bibinfo{author}{\bibfnamefont{S.~T.} \bibnamefont{West}} \bibnamefont{and}
  \bibinfo{author}{\bibfnamefont{J.~F.} \bibnamefont{Wheater}},
  \bibinfo{journal}{Nucl. Phys.} \textbf{\bibinfo{volume}{B486}},
  \bibinfo{pages}{261} (\bibinfo{year}{1997})

\bibitem[{\citenamefont{Boorstein and Kutasov}(1995)}]{Boorstein:1994rc}
\bibinfo{author}{\bibfnamefont{J.}~\bibnamefont{Boorstein}} \bibnamefont{and}
  \bibinfo{author}{\bibfnamefont{D.}~\bibnamefont{Kutasov}},
  \bibinfo{journal}{Phys. Rev.} \textbf{\bibinfo{volume}{D51}},
  \bibinfo{pages}{7111} (\bibinfo{year}{1995})

\bibitem[{\citenamefont{Polyakov}(1978)}]{Polyakov:1978vu}
\bibinfo{author}{\bibfnamefont{A.~M.} \bibnamefont{Polyakov}},
  \bibinfo{journal}{Phys. Lett.} \textbf{\bibinfo{volume}{B72}},
  \bibinfo{pages}{477} (\bibinfo{year}{1978}).

\bibitem[{\citenamefont{McLerran and Svetitsky}(1981)}]{McLerran:1981pb}
\bibinfo{author}{\bibfnamefont{L.~D.} \bibnamefont{McLerran}} \bibnamefont{and}
  \bibinfo{author}{\bibfnamefont{B.}~\bibnamefont{Svetitsky}},
  \bibinfo{journal}{Phys. Rev.} \textbf{\bibinfo{volume}{D24}},
  \bibinfo{pages}{450} (\bibinfo{year}{1981}).

\bibitem[{\citenamefont{Gupta et~al.}(2010)\citenamefont{Gupta, Mohapatra,
  Srivastava, and Tiwari}}]{Gupta:2010pp}
\bibinfo{author}{\bibfnamefont{U.~S.} \bibnamefont{Gupta}},
  \bibinfo{author}{\bibfnamefont{R.~K.} \bibnamefont{Mohapatra}},
  \bibinfo{author}{\bibfnamefont{A.~M.} \bibnamefont{Srivastava}},
  \bibnamefont{and} \bibinfo{author}{\bibfnamefont{V.~K.}
  \bibnamefont{Tiwari}}, \bibinfo{journal}{Phys. Rev.}
  \textbf{\bibinfo{volume}{D82}}, \bibinfo{pages}{074020}
  (\bibinfo{year}{2010})

\bibitem[{\citenamefont{Gupta et~al.}(2012)\citenamefont{Gupta, Mohapatra,
  Srivastava, and Tiwari}}]{Gupta:2011ag}
\bibinfo{author}{\bibfnamefont{U.~S.} \bibnamefont{Gupta}},
  \bibinfo{author}{\bibfnamefont{R.~K.} \bibnamefont{Mohapatra}},
  \bibinfo{author}{\bibfnamefont{A.~M.} \bibnamefont{Srivastava}},
  \bibnamefont{and} \bibinfo{author}{\bibfnamefont{V.~K.}
  \bibnamefont{Tiwari}}, \bibinfo{journal}{Phys.Rev.}
  \textbf{\bibinfo{volume}{D86}}, \bibinfo{pages}{125016}
  (\bibinfo{year}{2012})

\bibitem[{\citenamefont{Layek et~al.}(2005)\citenamefont{Layek, Mishra, and
  Srivastava}}]{Layek:2005fn}
\bibinfo{author}{\bibfnamefont{B.}~\bibnamefont{Layek}},
  \bibinfo{author}{\bibfnamefont{A.~P.} \bibnamefont{Mishra}},
  \bibnamefont{and} \bibinfo{author}{\bibfnamefont{A.~M.}
  \bibnamefont{Srivastava}}, \bibinfo{journal}{Phys. Rev.}
  \textbf{\bibinfo{volume}{D71}}, \bibinfo{pages}{074015}
  (\bibinfo{year}{2005})

\bibitem{kbl} T.W.B. Kibble, {\it J. Phys. A} {\bf 9}, 1387 (1976);
Phys. Rept. {\bf 67}, 183 (1980).

\bibitem{qurk1} V.M. Belyaev, Ian I. Kogan, G.W. Semenoff, and
N. Weiss, Phys. Lett. {\bf B277}, 331 (1992), A. V. Smilga,
Ann. Phys. {\bf 234},1 (1994).

\bibitem{qurk2}  C.P. Korthals Altes,  hep-th/9402028

\bibitem{psrsk} R.D. Pisarski, Phys. Rev. {\bf D62}, 111501R (2000);
{\it ibid}, hep-ph/0101168.

\bibitem{psrsk2} A. Dumitru and R.D. Pisarski, Phys. Lett. {\bf B 504},
282 (2001); Phys. Rev. {\bf D 66}, 096003 (2002); Nucl. Phys.
{\bf A698}, 444 (2002).

\bibitem[{\citenamefont{Deka et~al.}(2010)\citenamefont{Deka, Digal, and
  Mishra}}]{Deka:2010bc}
\bibinfo{author}{\bibfnamefont{M.}~\bibnamefont{Deka}},
  \bibinfo{author}{\bibfnamefont{S.}~\bibnamefont{Digal}}, \bibnamefont{and}
  \bibinfo{author}{\bibfnamefont{A.~P.} \bibnamefont{Mishra}}, 
  \bibinfo{journal}{Phys. Rev.} \textbf{\bibinfo{volume}{D85}}, 
  \bibinfo{pages}{114505} (\bibinfo{year}{2012})

\bibitem[{\citenamefont{Layek et~al.}(2006)\citenamefont{Layek, Mishra,
  Srivastava, and Tiwari}}]{Layek:2005zu}
\bibinfo{author}{\bibfnamefont{B.}~\bibnamefont{Layek}},
  \bibinfo{author}{\bibfnamefont{A.~P.} \bibnamefont{Mishra}},
  \bibinfo{author}{\bibfnamefont{A.~M.} \bibnamefont{Srivastava}},
  \bibnamefont{and} \bibinfo{author}{\bibfnamefont{V.~K.}
  \bibnamefont{Tiwari}}, \bibinfo{journal}{Phys. Rev.}
  \textbf{\bibinfo{volume}{D73}}, \bibinfo{pages}{103514}
  (\bibinfo{year}{2006})

\bibitem[{\citenamefont{Korthals~Altes and
  Watson}(1995)}]{KorthalsAltes:1994if}
\bibinfo{author}{\bibfnamefont{C.~P.} \bibnamefont{Korthals~Altes}}
  \bibnamefont{and} \bibinfo{author}{\bibfnamefont{N.~J.}
  \bibnamefont{Watson}}, \bibinfo{journal}{Phys. Rev. Lett.}
  \textbf{\bibinfo{volume}{75}}, \bibinfo{pages}{2799} (\bibinfo{year}{1995})

\bibitem[{\citenamefont{Atreya et~al.}(2012)\citenamefont{Atreya, Srivastava,
  and Sarkar}}]{Atreya:2011wn}
\bibinfo{author}{\bibfnamefont{A.}~\bibnamefont{Atreya}},
  \bibinfo{author}{\bibfnamefont{A.~M.} \bibnamefont{Srivastava}},
  \bibnamefont{and} \bibinfo{author}{\bibfnamefont{A.}~\bibnamefont{Sarkar}},
  \bibinfo{journal}{Phys.Rev.} \textbf{\bibinfo{volume}{D85}},
  \bibinfo{pages}{014009} (\bibinfo{year}{2012})

\bibitem[{\citenamefont{Korthals~Altes
  et~al.}(1994)\citenamefont{Korthals~Altes, Lee, and
 Pisarski}}]{KorthalsAltes:1994be}
 \bibinfo{author}{\bibfnamefont{C.~P.} \bibnamefont{Korthals~Altes}},
 \bibinfo{author}{\bibfnamefont{K.-M.} \bibnamefont{Lee}}, \bibnamefont{and}
 \bibinfo{author}{\bibfnamefont{R.~D.} \bibnamefont{Pisarski}},
 \bibinfo{journal}{Phys.Rev.Lett.} \textbf{\bibinfo{volume}{73}},
 \bibinfo{pages}{1754} (\bibinfo{year}{1994})

\bibitem[{\citenamefont{Korthals~Altes}(1992)}]{KorthalsAltes:1992us}
  \bibinfo{author}{\bibfnamefont{C.~P.} \bibnamefont{Korthals~Altes}}
  (\bibinfo{year}{1992}), \bibinfo{note}{in *Dallas 1992, Proceedings, High
  energy physics, vol. 2* 1443-1447}.

\bibitem{satz} H. Satz, J.Phys. {\bf G32}, R25 (2006).

\bibitem[{\citenamefont{Giannuzzi and Mannarelli}(2009)}]{Giannuzzi:2009gb}
\bibinfo{author}{\bibfnamefont{F.}~\bibnamefont{Giannuzzi}} \bibnamefont{and}
  \bibinfo{author}{\bibfnamefont{M.}~\bibnamefont{Mannarelli}},
  \bibinfo{journal}{Phys.Rev.} \textbf{\bibinfo{volume}{D80}},
  \bibinfo{pages}{054004} (\bibinfo{year}{2009})

\bibitem{csmrscl} G. Lacroix, C. Semay, D. Cabrera, and F. Buisseret
Phys. Rev. {\bf D87}, 054025 (2013);
G. S. Bali, Phys. Rev. {\bf D 62}, 114503 (2000),
see also, N. Cardoso, M. Cardoso, P. Bicudo, arXiv:1108.1542 (2011).

\bibitem{octerearly} A. Nakamura and T. Saito, Phys. Lett. {\bf B 621},
171 (2005).

\bibitem{octetnew} Y. Nakagawa, A. Nakamura, T. Saito, and H. Toki,
Phys. Rev. {\bf D77}, 034015 (2008).

\bibitem[{\citenamefont{Digal et~al.}(2005)\citenamefont{Digal, Kaczmarek,
  Karsch, and Satz}}]{Digal:2005ht}
\bibinfo{author}{\bibfnamefont{S.}~\bibnamefont{Digal}},
  \bibinfo{author}{\bibfnamefont{O.}~\bibnamefont{Kaczmarek}},
  \bibinfo{author}{\bibfnamefont{F.}~\bibnamefont{Karsch}}, \bibnamefont{and}
  \bibinfo{author}{\bibfnamefont{H.}~\bibnamefont{Satz}},
  \bibinfo{journal}{Eur.Phys.J.} \textbf{\bibinfo{volume}{C43}},
  \bibinfo{pages}{71} (\bibinfo{year}{2005})

\end{thebibliography}

\end{document}